# G-NetMon: A GPU-accelerated Network Performance Monitoring System for Large Scale Scientific Collaborations

Wenji Wu, Phil DeMar, Don Holmgren, Amitoj Singh, Ruth Pordes
Computing Division, Fermilab
Batavia, IL 60510, USA
E-mail: {wenji, demar, djholm, amitoj, ruth}@fnal.gov

*Abstract*—Network traffic is difficult to monitor and analyze, especially in high-bandwidth networks. Performance analysis, in particular, presents extreme complexity and scalability challenges. GPU (Graphics Processing Unit) technology has been utilized recently to accelerate general purpose scientific and engineering computing. GPUs offer extreme thread-level parallelism with hundreds of simple cores. Their data-parallel execution model can rapidly solve large problems with inherent data parallelism. At Fermilab, we have prototyped a GPU-accelerated network performance monitoring system, called G-NetMon, to support large-scale scientific collaborations. In this work, we explore new opportunities in network traffic monitoring and analysis with GPUs. Our system exploits the data parallelism that exists within network flow data to provide fast analysis of bulk data movement between Fermilab and collaboration sites. Experiments demonstrate that our G-NetMon can rapidly detect sub-optimal bulk data movements.

*Keywords: GPU, Flow Analysis, Network Performance Monitoring, High-speed netwworks.*

## I. Introduction

Large-scale research efforts such as Large Hadron Collider experiments and climate modeling are built upon large, globally distributed collaborations. The datasets associated with these projects commonly reach petabytes or tens of petabytes per year. The ability to efficiently retrieve, store, analyze, and redistribute the datasets generated by these projects is extremely challenging. Such projects depend on predictable and efficient data transfers between collaboration sites. However, achieving and sustaining efficient data movement over high-speed networks with TCP remains an on-going challenge. Obstacles to efficient and sustainable data movement arise from many causes and can create major impediments to the success of large-scale science collaborations. In practice, most sub-optimal data movement problems go unnoticed. Ironically, although various performance debugging tools and services are available to assist in identifying and locating performance bottlenecks, these tools cannot be applied until a problem is detected. In many cases, effective measures are not taken to fix a performance problem simply because the problem is either not detected at all or not detected in a timely manner. Therefore, it is extremely beneficial to possess a set of tools or services that can quickly detect sub-optimal data movement for large-scale scientific collaborations.

Generally speaking, network traffic is difficult to monitor and analyze. Existing tools like Ping, Traceroute, OWAMP [1] and SNMP provide only coarse-grained monitoring and diagnosis data about network status [2][3]. It is very difficult to use these tools to detect sub-optimal data movement. For example, SNMP-based monitoring systems typically provide 1-minute or 5-minute averages for network performance data of interest. These averages may obscure the instantaneous network status. On the other extreme, packet trace analysis [4][5] involves traffic scrutiny on a per-packet basis and requires high-performance computation and large-volume storage. It faces extreme scalability challenges in high-speed networks, especially as network technology evolves towards 100 Gbps. Flow-based data analysis, using router-generated flow-data such as Cisco's NetFlow [6] lies in between the two extremes. It produces a finer-grained analysis than SNMP, yet much less complex or voluminous as packet trace analysis. In this paper, we use flow-based analysis to detect sub-optimal data movements for large-scale scientific collaborations.

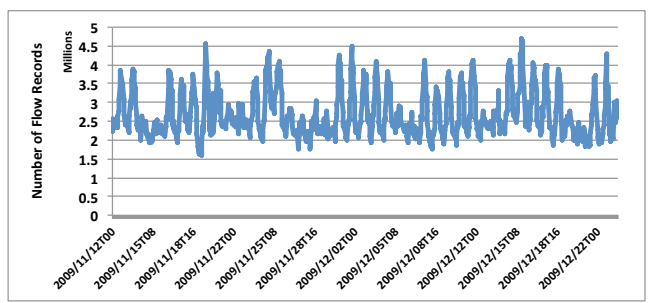

Figure 1 Number of Flow Records Generated at Fermilab Border Routers

To quickly detect sub-optimal data movements, it is necessary to calculate transfer rates between collaboration sites on an ongoing basis. Sub-optimal bulk data movement is detected if the associated transfer rate falls below some standard that is either predefined or provided by other network services. To this end, we use network flow data to calculate transfer rates between Fermilab and collaboration sites. Our flow-based analysis requires traffic scrutiny on a per-flow-record basis. In high-bandwidth networks, hundreds of thousands of flow records are generated each minute. Fermilab is the Tier-1 Center for the Large Hadron Collider's (LHC) Compact Muon Solenoid (CMS) experiment, as well as the central data center for several other large-scale research collaborations. Scientific data (e.g., CMS) dominates off-site traffic volumes in both inbound and outbound directions. Every hour, millions of flow records are generated at Fermilab border routers (Figure 1). Processing that much flow data in near real time requires both enormous raw compute power and high I/O throughputs.

Recently, GPU technology has been employed to accelerate general purpose scientific and engineering computing. GPUs offer extreme thread-level parallelism with hundreds of simple cores. The massive array of GPU cores offers an order of magnitude higher raw computation power than a conventional CPU. Its data-parallel execution model and ample memory bandwidth effectively hide memory access latency and can boost I/O intensive applications with inherent data parallelism.

At Fermilab, we have prototyped a GPU-accelerated network performance monitoring system (G-NetMon) for our large-scale scientific collaborations. In this work, we explore new opportunities in network traffic monitoring and analysis with GPUs. G-NetMon exploits the inherent data parallelism that exists within network flow data and uses a GPU to rapidly calculate transfer rates between Fermilab and collaboration sites in near real time. Experiments demonstrate that GPU can accelerate network flow data processing by a factor of 5 or more. G-NetMon can rapidly detect sub-optimal bulk data movement.

The rest of the paper is organized as follows. In section 2, we discuss some background and related work. In section 3, we introduce our G-NetMON design. In section 4, we discuss the experiments we used to evaluate how GPU can accelerate network flow data processing in high-speed network environments. Also, we evaluate how our system can effectively detect sub-optimal data transfer between Fermilab and collaboration sites. Finally, Section 5 concludes the paper.

## II. BACKGROUND & RELATED WORK

The rapidly growing popularity of GPUs makes them a natural choice for high-performance computing. Our GPU-accelerated network performance monitoring system is based on NVIDIA's Tesla C2070, featuring NVIDIA's latest Fermi GPU architecture. In the following sections, we give a simple introduction of NVIDIA's CUDA programming model and the Fermi GPU architecture.

### A. CUDA and the Fermi GPU Architecture

CUDA is the hardware and software architecture that enables NVIDIA GPUs to execute programs written with C, C++, and other languages. It provides a simple programming model that allows application developers to easily program GPU and explicitly express parallelism. A CUDA program consists of parts that are executed on the host (CPU) and parts on the GPU. The parts that exhibit little or no data parallelism are implemented as sequential CPU threads. The parts that exhibit a rich amount of data parallelism are implemented as GPU kernels. GPU instantiates a kernel program on a grid of parallel thread blocks. Each thread within a thread block executes an instance of the kernel, and has a per-thread ID, program counter, registers, and private memory. Threads within a thread block can cooperate among themselves through barrier synchronization and shared memory. Thread blocks are grouped into grids, each of which executes a unique kernel. Each thread block has a unique block ID. A thread indexes its data with its respective thread ID and block ID.

NVIDIA's Fermi GPU architecture consists of multiple streaming multiprocessors (SMs), each consisting of 32 CUDA cores. A CUDA core executes a floating-point or integer instruction per clock for a thread. Each SM has 16 load/store units, allowing source and destination addresses to be calculated for sixteen threads per clock and 4 special function units (SFUs) to execute transcendental instructions. The SM schedules threads in groups of 32 parallel threads called warps. Each SM features two warp schedulers and two instruction dispatch units, allowing two warps to be issued and executed concurrently. The execution resources in a SM include registers, thread block slots, and thread slots. These resources are dynamically partitioned and assigned to threads to support their execution. We list these resource limits per SM in Table 1. In addition, the Fermi GPU has six 64-bit memory partitions, for a 384-bit memory interface, supporting up to a total of 6 GB of GDDR5 DRAM memory. A host interface connects the GPU to the CPU via PCI-Express. The GigaThread global scheduler distributes thread blocks to SM thread schedulers.

**Table 1 Physical Limits per SM for Fermi GPU**

| | |
|---|---|
| Maximum Warps: | 48 |
| Maximum Threads: | 1536 |
| Maximum Blocks: | 8 |
| Shared Memory: | 48K |
| Register Count: | 32K |

### B. GPU in Network Related Applications

GPU offers extreme thread-level parallelism with hundreds of simple cores. The massive array of GPU cores offers an order of magnitude higher raw computation power than a conventional CPU. GPU's data-parallel execution model and ample memory bandwidth fits nicely with most networking applications, which have inherent data parallelism at either packet level or at network data flow level. Recently, GPUs have shown a substantial performance boost to many network applications, including GPU-accelerated software router [7], pattern matching [8][9][10], network coding [11], IP table lookup [8], and cryptography [12]. So far, the application of GPU in network applications is manly focusing at packet level. In this work, we make use of GPU to accelerate network flow data analysis.

### C. Flow-based Analysis

Flow-based analysis is widely used in traffic engineering [13][14], anomaly detection [15][16], traffic classification [17][18], performance analysis, and security [19][20][21], etc. For example, Internet2 makes use of flow data to generate traffic summary information by breaking the data down in a number of ways, including by IP protocol, by a well-known service or application, by IP prefixes associated with "local" networks, or by the AS pairs between which the traffic was exchanged. In [15], the sub-space method is applied to flow traffic to detect network-wide anomalies.

### III. G-NETMON SYSTEM DESIGN

To quickly detect sub-optimal data movements, G-NetMon uses network flow data to calculate transfer rates between Fermilab and collaboration sites on an on-going basis. A sub-optimal bulk data movement is detected if the associated transfer rates fall below some standard that is either predefined or provided by other network services. Our GPU-accelerated network performance monitoring system is deployed as shown in Figure 2. It receives flow data from site border routers as well as internal LAN routers. The routers export NetFlow V5 records. The flow data is complete, not sampled.

*A. System Hardware Configuraton*

Our flow-based analysis requires traffic scrutiny on a per-flow-record basis. Fermilab is the US-CMS Tier-1 Center and the main data center for a few other large-scale research collaborations. Every hour, millions of flow records are generated at Fermilab border routers (Figure 1). Considering the increasing volume of scientific data created every year, coupled with the evolution towards to 100 GigE network technologies, it is anticipated that our network flow data analysis requirements will be increasing accordingly. Therefore, our G-NetMon not only needs to handle current network conditions, but have the capability to accommodate the large growth of traffic expected in the near future. For now, Fermilab border routers generate less than 5,000,000 flow records every hour. Our target is to allow G-NetMon to handle 50,000,000 flow records per hour.

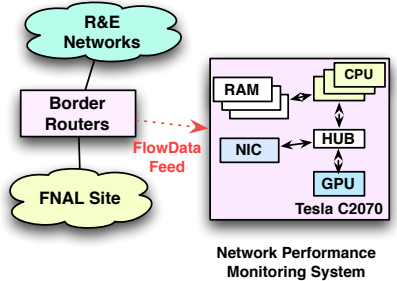

**Figure 2 G-NetMon – Deployment**

G-NetMon is implemented in a system that consists of two 8-Core 2.4 GHz AMD Opteron 6136 processors, two 1Gbps Ethernet interfaces, 32 GB of system memory, and one Tesla C2070 GPU. The Tesla C2070 GPU features the Fermi GPU architecture. Its key features are listed in Table 2.

**Table 2 Tesla C2070 Key Features**

| | |
|---|---|
| SMs: | 14 |
| Cores: | 448 |
| Core Freq.: | 1.15 GHz |
| Global Memory Size: | 6GB GDDR5 |
| Memory Bandwidth: | 144 GB/s |
| System Interface: | PCIex16 Gen2 |
| Double Precision Peak Performance: | 515 GFlops |

*B. System Architecture*

The G-NetMon architecture is as shown in Figure 3. The system consists of a few parts that are executed on either the host (CPU) or GPU. Based on the CUDA design principle, the parts that exhibit little or no data parallelism are implemented as sequential CPU threads; the parts that exhibit a rich amount of data parallelism are implemented as GPU kernels.

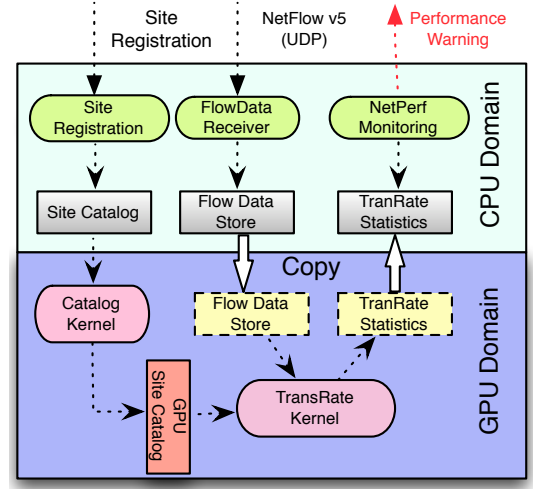

**Figure 3 A G-NetMon – Architecture**

*B.1 CPU Domain*

Three CPU threads are implemented in the CPU domain.

**Site Registration Thread:** it registers scientific subnets to our network performance monitoring system. The registered subnets are stored in the Site Catalog (a data buffer in host memory), which helps to identify scientific data transfer between Fermilab and collaboration sites. Large-scale research efforts like LHC CMS are built upon large, globally distributed collaborations. However, available computing and networking resources at different collaboration sites varies greatly. Larger sites, such as Fermilab, have data centers comprising thousands of computation nodes that function as massively scaled, highly distributed cluster-computing platforms. These sites are usually well connected to the outside world with high-bandwidth links of 10 Gbps or more. On the other hand, some small collaboration sites have limited computing resources and significantly lower bandwidth-networking connectivity. Therefore, scientific data transfers between collaboration sites can vary greatly in terms of performance and scale. It is difficult to design machine-learning algorithms to automatically identify scientific data transfers in terms of traffic patterns or characteristics. However, for a large-scale scientific application, the collaboration relationships between research institutions tend to be relatively static. In addition, the systems and networks assigned to a scientific application at a site are relatively fixed. Large-scale scientific data movement usually occurs between some specific subnets at each site. Therefore, by registering these subnets to our system, we can easily monitor data transfers between Fermilab and its collaboration sites through flow analysis of traffic between those subnets.

**FlowData Receiver Thread:** a UDP daemon, which receives NetFlow V5 packets from border routers. The received flow records are stored in Flow Data Store (a data buffer in host memory). In the current implementation, Flow Data Store is designed to hold 50,000,000 flow records. Since a NetFlow V5 flow record is less than 50 Bytes, these 50,000,000 flow records require approximately 2.5GB of memory. Processed flow records in Flow Data Store are periodically cleaned and stored to disk to create space for subsequent network flow data.

**NetPerf Monitoring Thread:** the main thread of our network performance monitoring system. Periodically (each hour), it copies Site Catalog and Flow Data Store to GPU memory and launches the corresponding GPU kernels to calculate the transfer rates between Fermilab and its collaboration sites. When GPU computation is completed, the NetPerf Monitoring Thread will synthesize the final results. A sub-optimal bulk data movement is detected if the associated transfer rates are below some predefined standard. Considering that TCP traffic is elastic, we use the statistics of transfer rate medians as our evaluation criteria. For a given site, network performance warnings would be issued if the associated median were less than 1Mbps for two consecutive hours.

*B.2 GPU Domain*

*1) GPU Kernels*

In the GPU domain, we have implemented two GPU kernels, Catalog Kernel and TransRate Kernel.

**Catalog Kernel:** it builds GPU Site Catalog, a hash table for registered scientific subnets in GPU memory, from Site Catalog. TransRate Kernel makes use of GPU Site Catalog to rapidly assign flow records to their respective subnets by examining their source or destination IP addresses. To make the hash table easy to implement and fast to search, all registered networks are transformed into /24 subnets and then entered in GPU Site Catalog. For the sake of scientific data transfer, a /24 subnet is large enough for most collaboration sites. Any network larger than /24 is divided into multiple entries in the hash table. Since GPU Site Catalog is mainly used for lookup operations and is rarely updated, there is no need to implement locks to protect unsynchronized write accesses. If any update is necessary, the table is rebuilt from scratch.

**TransRate Kernel:** it calculates the transfer rates between Fermilab and its collaboration sites. TransRate Kernel exploits the inherent data parallelism that exists within network flow data. When GPU instantiates TransRate Kernel on a grid of parallel threads, each thread handles a separate flow record. On a C2070 GPU, thousands of flow records can be processed simultaneously. To handle a flow record, a TransRate thread first attempts to assign the flow record to its respective site and then calculates the corresponding transfer rates. With a hash of the /24 subnet of the flow record's source or destination IP address, TransRate Kernel looks up the site to which the flow record belongs in GPU Site Catalog. Because each flow record includes data such as the number of packets and bytes in the flow and the timestamps of the first and last packet, calculation of transfer rate is simple. However, two additional factors must be considered. First, because a TCP connection is bidirectional, it will generate two flow records, one in each direction. In practice, a bulk data movement is usually unidirectional. Only the flow records in the forward direction reflect the true data transfer activities. The flow records in the other direction simply record the pure ACKs of the reverse path and should be excluded from transfer rate calculations. These flow records can be easily filtered out by calculating their average packet size, which is usually small. Second, a bulk data movement usually involves frequent administrative message exchanges between the two endpoints. A significant number of flow records are generated due to these activities. These records usually contain a small number of packets with short durations; their calculated transfer rates are generally of low accuracy and high variability. These flow records are also excluded from our transfer rate calculation.

We calculate transfer rates (maximum, minimum, average, median) for each registered site and for each host in a registered site. To calculate the median statistics, we create an array of buckets for each host to count transfer rate frequencies. Each bucket represents a 10kbps interval. To save space, all transfer rates greater than 100Mpbs are counted in the last bucket. Therefore, for each host, we maintain a bucket array of size 10001. A bucket *n* represents the frequency of flow rates that fall within the interval [n*10kbps (n+1)*10kbps]. From the resulting bucket counts we determine the host and site medians. We use atomic CUDA operations to calculate and store all transfer rates in order to prevent unsynchronized data accesses by the threads.

*2). GPU Kernel Optimization*

The Catalog Kernel is relatively simple, with few opportunities for optimization. In fact, its functionality could be included in TransRate Kernel. However, because the overhead to launch a kernel is negligible [7], we have chosen to implement it as an independent kernel to preserve a modular design.

Our TransRate kernel is optimized using various approaches:

- Register Spilling Optimization. Without this optimization, a TransRate thread will use 47 registers. These registers hold compiler-generated variables. Because registers are in-chip memories that can be accessed rapidly, a single thread's performance increases if registers are readily available. However, when we used the CUDA Occupancy Calculator to measure SM occupancy with varying block sizes, to our surprise, the occupancy rates were unacceptably low (Table 3). At such a low SM occupancy, the overall GPU performance would be greatly degraded. The improvement in each single thread cannot make up for the loss in overall thread parallelism. To raise GPU occupancy, we limit the maximum number of registers used by TransRate to 20 by compiling this kernel with the "-maxrregcount 20" option. As shown in Table 3, this register spilling optimization is effective, and the best GPU occupancy achieved as the number of threads per block is varied is now 100%.
- Shared memory. Shared memories are on-chip memories and can be accessed at very high speed in a highly parallel manner. The TransRate kernel makes use of shared memory as much as possible to accelerate flow data processing.

- Non-caching Load. Fermi architecture global memory has two types of loads, caching and non-caching. The caching load is the default mode. It first attempts to load from L1 cache, then from L2 cache, and finally from the global memory. The load granularity is 128 bytes. The non-caching load first attempts to hit in L2, and then the global memory. Its load granularity is 32 bytes. Our experiments show that non-caching load can boost the performance by at least 10%, and so the optimized TransRate kernel uses non-caching load to access Flow Data Store.

**Table 3 SM Occupancy Rates at Different Kernel Block Sizes**

| Thread Size per Block | 64 | 128 | 256 | 512 |
|---|---|---|---|---|
| SM Occupancy Rates @ Register/Thread=47 | 33% | 42% | 33% | 33% |
| SM Occupancy Rates @ Register/Thread=20 | 33% | 67% | 100% | 100% |

## IV. EXPERIMENTAL EVALUATION

In this section, we show results of our experimental evaluation of G-NetMon. First, we evaluate the performance of our G-NetMon system. Also, we study how GPU can accelerate network flow data processing in high-volume network data flow environments. Second, we deploy our G-NetMon in Fermilab production environments. We evaluate how G-NetMon can effectively detect sub-optimal data transfers between Fermilab and its collaboration sites.

### A. Performance Evaluation

At present, Fermilab border routers produce fewer than 5,000,000 flow records in an hour. However, our G-NetMon system is designed to handle a maximum load of 50,000,000 flow records per hour. To evaluate the capabilities and performance of our system at such a network load, we collected more than a day's flow records from the border routers and fed G-NetMon with 50,000,000 flow records. FlowData Receiver Thread receives these flow records and stores them in Flow Data Store. We also select the top 100 /24 scientific subnets that transfer to and from Fermilab in terms of traffic volume, and register them with Site Catalog.

#### A.1 GPU Performance & Optimization

To evaluate GPU performance, and the effects of various GPU kernel optimization approaches, we have implemented several G-NetMon variants with different enabled optimizations. Our objectives are to compare effects:
- Shared-Memory vs. Non-Shared-Memory. For Non-Shared-Memory, the TransRate Kernel does not use shared memory, and all the operations are executed on GPU global memory.
- Caching-Load vs. Non-Caching-Load.
- Hash-Table-Search vs. Non-Hash-Table-Search (sequential search). To calculate transfer rates between Fermilab and collaboration sites, it is first necessary to assign flow records to their respective sites. G-NetMon implements a hash table to perform this function. We have also implemented a Non-Hash-Table method (i.e., sequential search) in which all of the registered scientific subnets are maintained in a sequential list. To categorize a flow record, the TransRate kernel searches the list one by one until a matching site, or none, is found.

We list all the G-NetMon variants according to enabled optimizations in Table 4. In the table, "Y" indicates that the "xxx" optimization is enabled, while "N" indicates the optimization is not used. We enabled the register-spilling optimization when compiling all of these GPU variants, and so the TransRate Kernel is launched with 100% occupancy. We ran experiments to measure the rates at which these G-NetMon variants process network flow data and compared them with the performance of the fully optimized G-NetMon.

**Table 4 GPU Variants with Different Features**

| GPU Variants | Features | | |
|---|---|---|---|
| | Hash Table | Share-Memory | Caching Load |
| G-NetMon | Y | Y | N |
| NH-S-C-GPU | N | Y | Y |
| NH-NS-C-GPU | N | N | Y |
| NH-NS-NC-GPU | N | N | N |
| H-NS-C-GPU | Y | N | Y |
| H-NS-NC-GPU | Y | N | N |
| H-S-C-GPU | Y | Y | Y |
| NH-S-NC-GPU | N | Y | N |

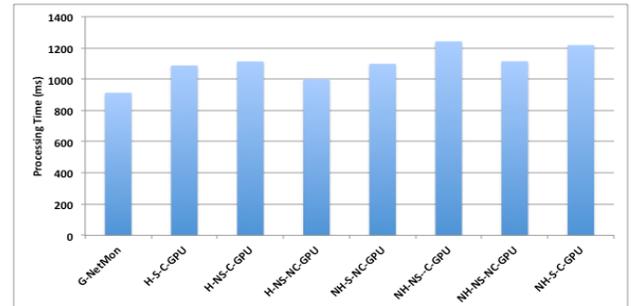

**Figure 4 GPU Processing Time**

The NetPerf Monitoring thread copies Site Catalog and Flow Data Store to GPU memory and launches the corresponding GPU kernels to calculate the transfer rates. We evaluate how fast GPU can handle these data. The experiment results are shown in Figure 4. To handle 50,000,000 flow records, G-NetMon takes approximately 900 milliseconds. The effects of the various GPU kernel optimizations are shown. For example, with the combination of hash table and non-caching-load, shared-memory can accelerate flow record processing by as much as 9.51%. As discussed above, shared memories are on-chip memories that can be accessed at very high speed in a highly parallel manner. The experiment results show that the hash table mechanism significantly boosts G-NetMon performance, ranging from 11% to 20%. We used NVIDIA's Compute Visual Profiler to profile the TransRate Kernel execution. Figure 5 gives the "Instruction Issued" comparisons of Hash-Table vs. Non-Hash-Table of all GPU code variants.

These experiments show that the hash table mechanism can significantly reduce flow record categorization overheads.

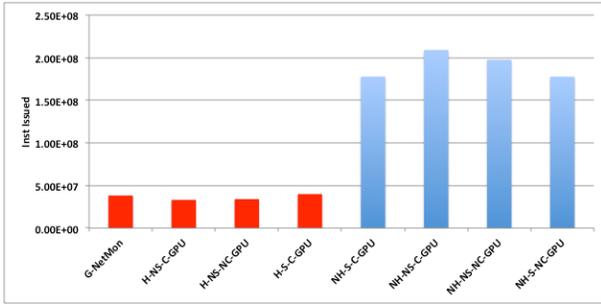

**Figure 5 Hash-Table (in Red) vs. Non-Hash-Table (in Blue)**

To our surprise, Figure 4 shows that non-caching-load boosts the performance significantly, by more than 10%. We speculate that the non-caching-load mode better fits G-NetMon's traffic pattern. When using the caching load, the performance gain in L1 cache does not compensate for the performance loss caused by the larger load granularity. Table 5 gives the comparisons of caching load vs. non-caching load for various memory access parameters. We see that caching load causes higher memory traffic, degrading the overall performance.

**Table 5 Caching-Load vs. Non-Caching-Load**

|  | **G-NetMon** | **H-S-C-GPU** |
|---|---|---|
| **L2 Read Requests** | 1.61835e+08 | 2.98727e+08 |
| **L2 Write Requests** | 1.15432e+07 | 1.61657e+07 |
| **Global Memory Read Requests** | 2.17803e+08 | 2.48466e+08 |
| **Global Memory Write Requests** | 3.01409e+07 | 3.22455e+07 |

*A.2 GPU vs. CPU*

In order to evaluate how GPU can accelerate network flow data processing in high-bandwidth network environments, we compare G-NetMon with its corresponding CPU implementations. We implemented two CPU variants, which are termed H-CPU and NH-CPU, respectively. Like G-NetMon, H-CPU applies a hash table mechanism to rapidly assign flow records to their respective sites and then calculates the corresponding transfer rates. In contrast, NH-CPU implements a similar Non-Hash-Table method (sequential search) as NH-S-NC-GPU, in which all of the registered scientific subnets are maintained in a sequential list. To assign a flow record, CPU searches the list one by one until a matching site, or none, is found. We ran each of H-CPU and NH-CPU on a single 2.4 GHz AMD Opteron 6136 core, with the same set of data as used above. We make the comparisons of G-NetMon vs. H-CPU and NH-S-NC-GPU vs. NH-CPU. The results are shown in Figure 6. It takes H-CPU 4916.67 ms to handle 50,000,000 flow records; in contrast, G-NetMon requires 900ms. For the non-hash-table variants, NH-CPU and NH-S-NC-GPU take 36336.67 ms and 1098.23 ms, respectively. The comparisons clearly show that GPU can significantly accelerate the flow data processing, by a factor of 5.38 (G-NetMon vs. H-CPU), or by a factor of 33.08 (NH-S-NC-GPU vs. NH-CPU). The reason that we present the comparison of GPU vs. CPU for the non-hash-table implementations is because many network applications feature a similar sequential search computation pattern as our non-hash-table implementations. For example, a network security application needs to examine each packet or flow with security rules one by one. The experiment results show GPU can significantly accelerate the data processing.

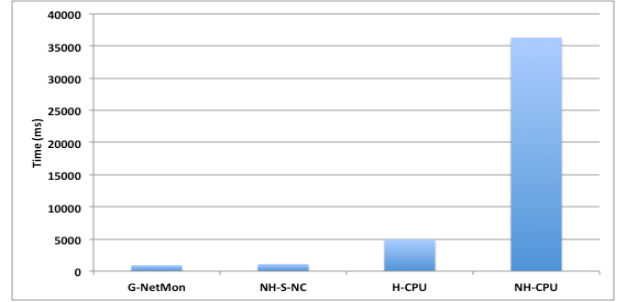

**Figure 6 G-NetMon vs. CPU**

*A.3 Receiving Flow Records*

G-NetMon receives NetFlow V5 packets from border routers via UDP. The received flow records are stored in Flow Data Store. A NetFlow V5 flow record is 48 bytes. A 1500-byte UDP packet, the largest allowed by standard Ethernet at the network, can transmit at most 30 flow records. Our G-NetMon system is designed to handle a maximum load of 50,000,000 flow records per hour. Therefore, the FlowData Receiver thread needs to handle at least 463 packets per second, which amounts to an average traffic load of 5.56Mbps. Our G-NetMon system can easily handle such a traffic load. However, because the flow records are transmitted via UDP, if CPU is busy with other tasks and the FlowData Receiver thread is not scheduled to handle the NetFlow traffic in time, the incoming packets can be dropped when the UDP receive buffer is full. We have run experiments to verify this scenario. In the experiments, the FlowData Receiver thread was assigned to share a core with a CPU-intensive application and the UDP receive buffer size was set to 4MB. We then sent it UDP traffic at varying rates, ranging from 100Mbps to 1Gbps, for 0.5 seconds. When the UDP traffic rates reached 500Mbps or above, serious packet loss would occur. We repeated the above experiments with the FlowData Receiver thread assigned a dedicated core. No packet loss was detected. Therefore, to avoid the situation of NetFlow packets being dropped, G-NetMon assigns a dedicated core for the FlowData Receiver thread to handle NetFlow traffic.

*B. Network Performance Monitoring*

We have registered 100 /24 scientific subnets that transfer to and from Fermilab in G-NetMon. G-NetMon monitors the bulk data movement status between Fermilab and these subnets by calculating the corresponding data transfer statistics every hour. G-NetMon calculates the transfer rates (maximum, minimum, average, median) for each registered site and for each host in a registered site. Figure 7 gives the data transfer rates in an hour between Fermilab and a collaboration site.

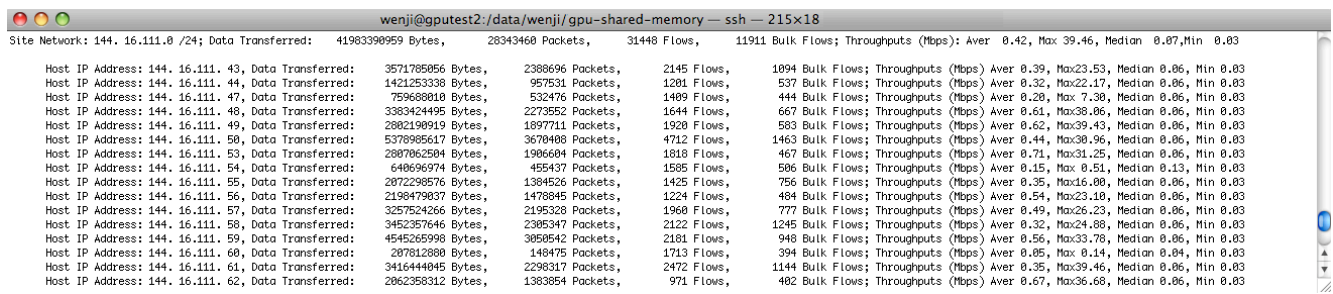

**Figure 7 Transfer rates between Fermilab and a Collaboration Site**

A sub-optimal bulk data movement is detected if the associated transfer rate falls below a predefined standard. Considering that TCP traffic is elastic and network conditions are volatile, we use the statistics of transfer rate medians as our evaluation criteria. For a given site, network performance warnings would be issued if the associated median were less than 1Mbps for two consecutive hours.

To evaluate the effectiveness of G-NetMon in detecting sub-optimal bulk data movements, we investigated the G-NetMon warnings for a period of two weeks. During this period, G-NetMon issued performance warnings for 7 sites in total (there were multiple warnings for the same sites). For those sites that G-NetMon issued warnings, we contacted their network administrators to conduct end-to-end performance analysis. Five sites responded to our requests. The end-to-end performance analysis indicated poor network conditions between these sites and Fermilab. To our surprise, one site in Greece is even connected to the outside world with a 100 Mbps link. The investigation of these warnings demonstrated that our G-NetMon can effectively detect sub-optimal bulk data movements in a timely manner. G-NetMon can detect a sub-optimal bulk data movement in two hours.

## V. CONCLUSION & DISCUSSION

At Fermilab, we have prototyped a GPU-accelerated network performance monitoring system for large-scale scientific collaborations, called G-NetMon. Our system exploits the inherent data parallelism that exists within network data flows and can rapidly analyze bulk data movements between Fermilab and its collaboration sites. Experiments demonstrate that our G-NetMon can detect sub-optimal bulk data movement in time.

Considering TCP traffic is elastic and network conditions are volatile, our G-NetMon system applies a very conservative approach to issue performance warnings. G-NetMon is chosen to perform transfer rate analysis every hour. Running G-NetMon with shorter intervals can detect sub-optimal bulk data movement faster. However, it would also generate more ephemeral warnings and finally degrade our system's effectiveness.

The main purpose of this work is to explore new opportunities in network traffic monitoring and analysis with GPUs. The experiment results show that GPU can significantly accelerate the flow data processing, by a factor of 5.38 (G-NetMon vs. H-CPU), or by a factor of 33.08 (NH-S-NC-GPU vs. NH-CPU). At present, G-NetMon is designed to detect sub-optimal bulk data movements. In the future, we will enhance it with security features. To implement security features, G-NetMon needs to examine flow records with security rules one by one in real time or semi-real time, which require more computation capabilities. The computation pattern of examining flow records with security rules one by one is similar to that of the non-hash-table implementations discussed in the paper, in which GPU can significantly accelerate the flow data processing.